\documentclass[graybox]{svmult}

\usepackage{mathptmx}       % selects Times Roman as basic font
\usepackage{helvet}         % selects Helvetica as sans-serif font
\usepackage{courier}        % selects Courier as typewriter font
\usepackage{makeidx}         % allows index generation
\usepackage{graphicx}        % standard LaTeX graphics tool
\usepackage{multicol}        % used for the two-column index
\makeindex             % used for the subject index
\begin{document}

\title*{Galaxy properties in different environments at $z > 1.5$ in the GOODS-NICMOS Survey}
\author{Ruth Gr\"utzbauch, Robert W. Chuter, Christopher J. Conselice, Amanda E. Bauer, Asa F. L. Bluck, Fernando Buitrago and Alice Mortlock}
\authorrunning{R. Gr\"utzbauch et al.}
\institute{all authors at University of Nottingham, \email{ruth.grutzbauch@nottingham.ac.uk}}
\maketitle

% Too much empty space in the original style file!
\vskip-1.2truein

\abstract{We present a study of the relationship between galaxy colour, stellar mass, and local galaxy density in a deep near-infrared imaging survey up to a redshift of $z\sim3$ using the GOODS NICMOS Survey (GNS). The GNS is a very deep, near-infrared Hubble Space Telescope survey imaging a total of 45 arcmin$^2$ in the GOODS fields, reaching a stellar mass completeness limit of $M_\ast = 10^{9.5}~M_\odot$ at $z=3$. Using this data we measure galaxy local densities based on galaxy counts within a fixed aperture, as well as the distance to the 3$^{rd}$, 5$^{th}$ and 7$^{th}$ nearest neighbour. We find a strong correlation between colour and stellar mass at all redshifts up to $z\sim3$.  We do not find a strong correlation between colour and local density, however, the highest overdensities might be populated by a higher fraction of blue galaxies than average or underdense areas, indicating a possible reversal of the colour-density relation at high redshift. Our data suggests that the possible higher blue fraction at extreme overdensities might be due to a lack of {\it massive} red galaxies at the highest local densities. 
}

\section{Introduction}
\label{sec:1}

Despite a wealth of recent studies, disentangling the influence of stellar mass and environment on galaxy formation and evolution remains a highly debated topic. The presence of a strong correlation between galaxy colour and stellar mass is supported by numerous observational studies in the local universe (see e.g. \cite{Kau03a,Kau03b}) and up to intermediate redshifts of $z\sim1$ (see e.g. \cite{Bun06,Gru10}).
However, galaxy colours not only depend on mass, but also on a galaxy's environment. The preference of red galaxies for denser local environments, first noticed by \cite{Oem74}, and confirmed by \cite{Dre80}, is now well-studied in the local universe (e.g. \cite{Kau04,vdW08}). In the early universe, however, the evidence is controversial. While some studies find that the colour-density relation at $z\sim1$ is mainly due to a bias in stellar mass selection and only persists for low-mass galaxies (e.g. \cite{Tas09,Iov10,Gru10}), others argue that a strong colour-density relation is already in place at $z\sim1.4$ (e.g. \cite{Coo06,Ger07}) even at fixed stellar mass \cite{Coo10}. In this study we investigating the relationship between galaxy colours, stellar mass and local densities in the critical redshift range at $z>1.5$, probing the period in which most of galaxy formation takes place. We use data from the GOODS NICMOS Survey (GNS), a large Hubble Space Telescope survey reaching a stellar mass completeness limit of $M_\ast \sim 10^{9.5}~M_\odot$ at $z\sim3$. 

\section{Data and Analysis}
\label{sec:2}

In this study we use data from the GOODS NICMOS Survey (GNS) covering a total area of about 45 arcmin$^2$ in the GOODS fields \cite{Con10}. The limiting magnitude reached at 5$\sigma$ is $H_{AB}$ = 26.8, which is more than 2 mag deeper than ground based near-infrared imaging \cite{Ret10}. To obtain photometric redshifts, stellar masses and rest-frame colours, the NICMOS $H$-band sources were matched to a catalogue of optical sources in the GOODS ACS fields, yielding a $BVizH$ photometric catalogue. 

The photometric redshifts are measured with the HYPERZ code \cite{Bol00}. For the high redshift sample ($z>1.5$) we obtain a photometric redshift uncertainty of $\sigma_{\Delta z /(1+z)} = 0.10$, while galaxies at $z<1.5$ show a slightly lower, but still comparable scatter of $\sigma_{\Delta z /(1+z)} = 0.08$. The photometric redshift uncertainties are used in a set of 100 Monte Carlo simulations to estimate the uncertainties in local density. For this purpose we randomize the photometric redshift input according to the photo-z error, accounting for scattering in and out of the redshift range, as well as dealing separately with catastrophic outliers. The randomized photo-zs are then used to recompute the local densities, giving typical average uncertainties of $\sim 0.2$ dex. We plot the results of the simulations in Figures~\ref{fig2} and \ref{fig3} discussed in Section~\ref{col-dens}.

The stellar masses and $(U-B)$ rest-frame colours are measured by fitting a large set of synthetic spectra constructed from the stellar population models of \cite{Bru03}, assuming a Salpeter initial mass function. The average uncertainty of our stellar mass estimate is $\sim 0.3$ dex.

We compute local densities using the fixed aperture method as well as the $N^{th}$ nearest neighbour approach. The absolute densities are corrected for edge effects and normalized by the median value in a redshift bin of $\Delta z = 0.25$, yielding a relative local density $\log~(1+\delta_N)$. Comparing the different density estimators we conclude that the $3^{rd}$ nearest neighbour density has the largest dynamical range and only slightly higher uncertainties and is therefore best suited to trace the extremes in the local density distribution. We show the results for $\log~(1+\delta_3)$ in the following.

\section{Results}
\label{subsec:2}

\subsection{Colour-magnitude relation}

\begin{figure}[t]
\begin{center}
\includegraphics[width=0.28\textwidth]{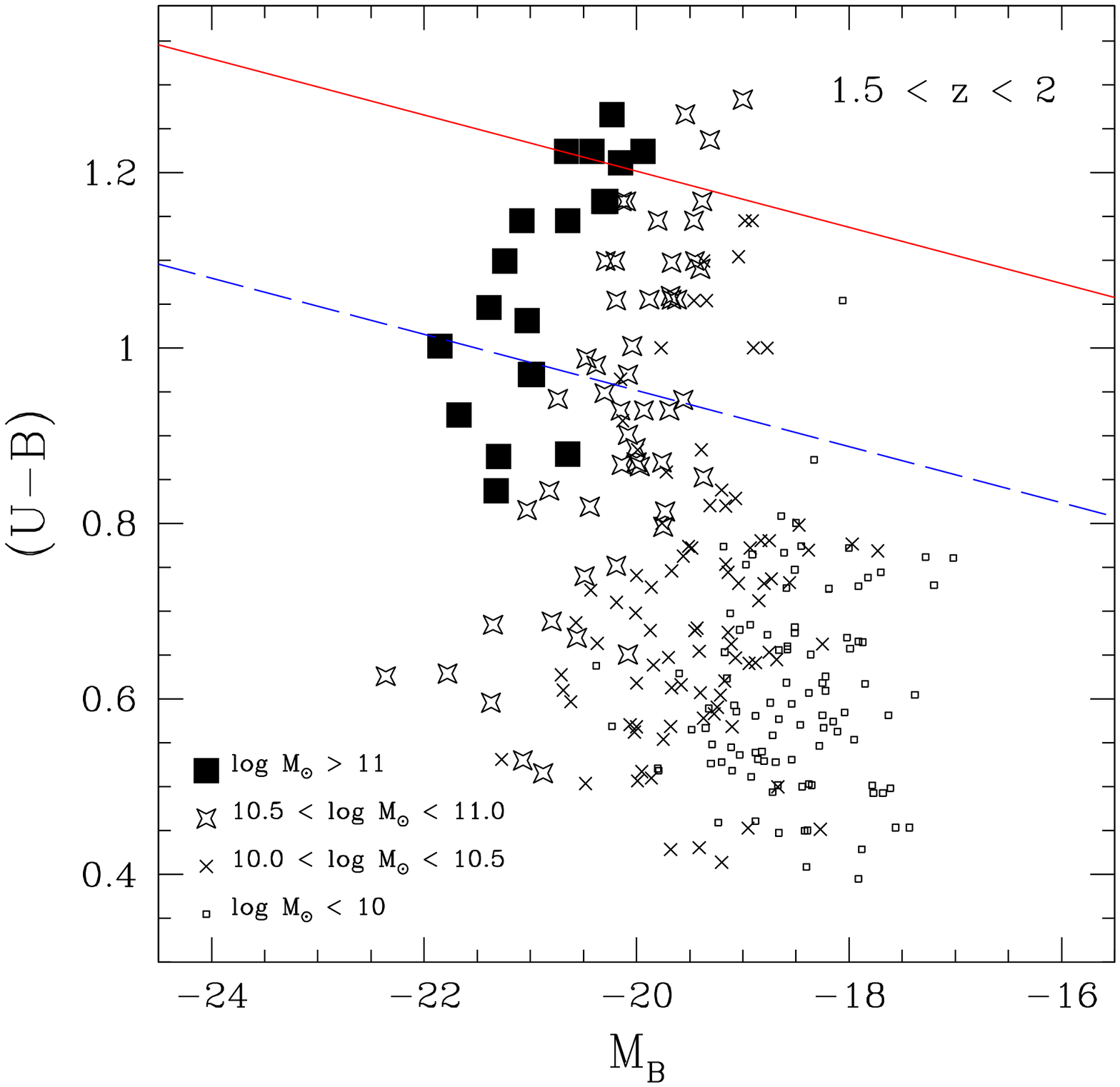}
\includegraphics[width=0.28\textwidth]{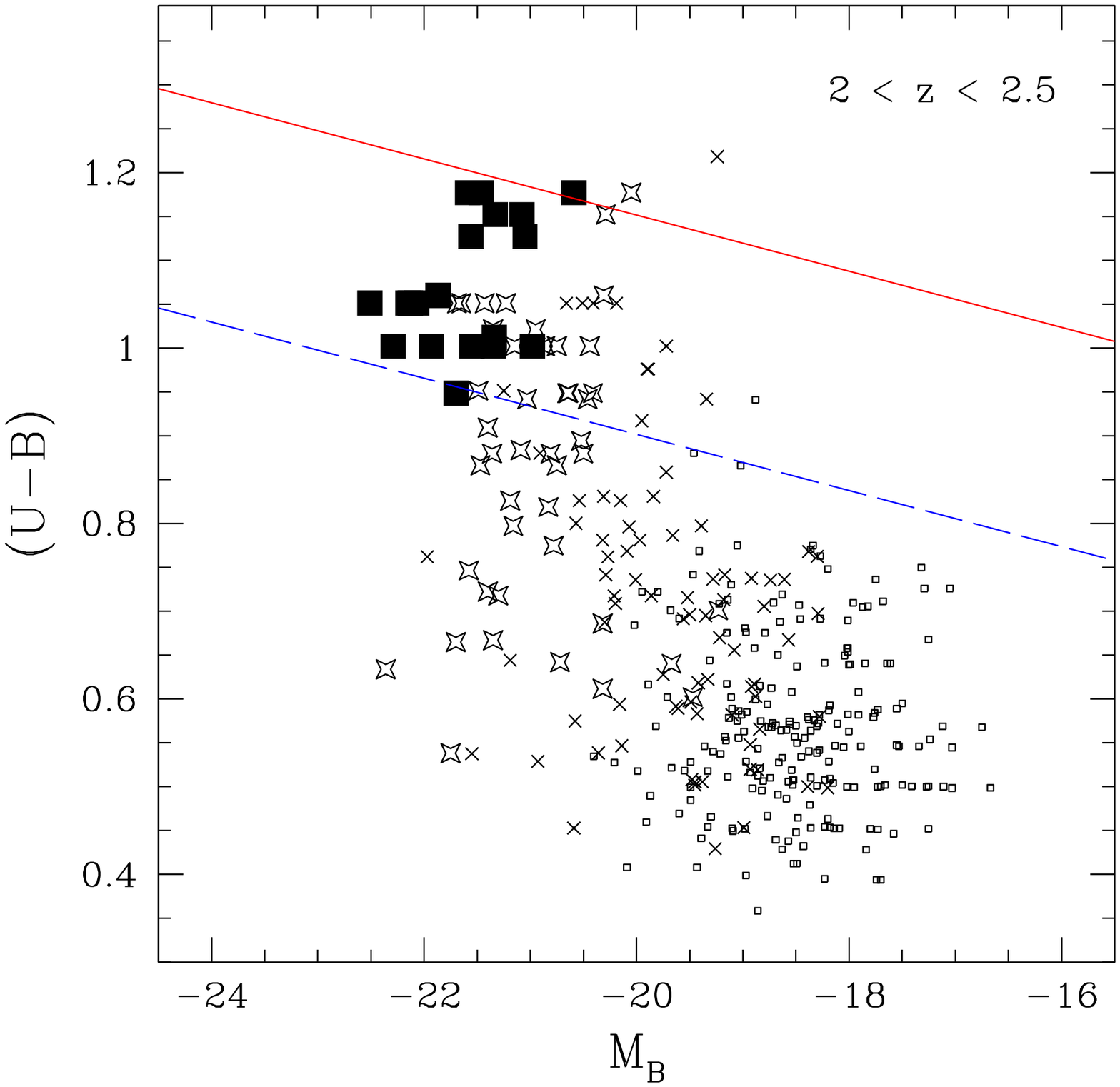}
\includegraphics[width=0.28\textwidth]{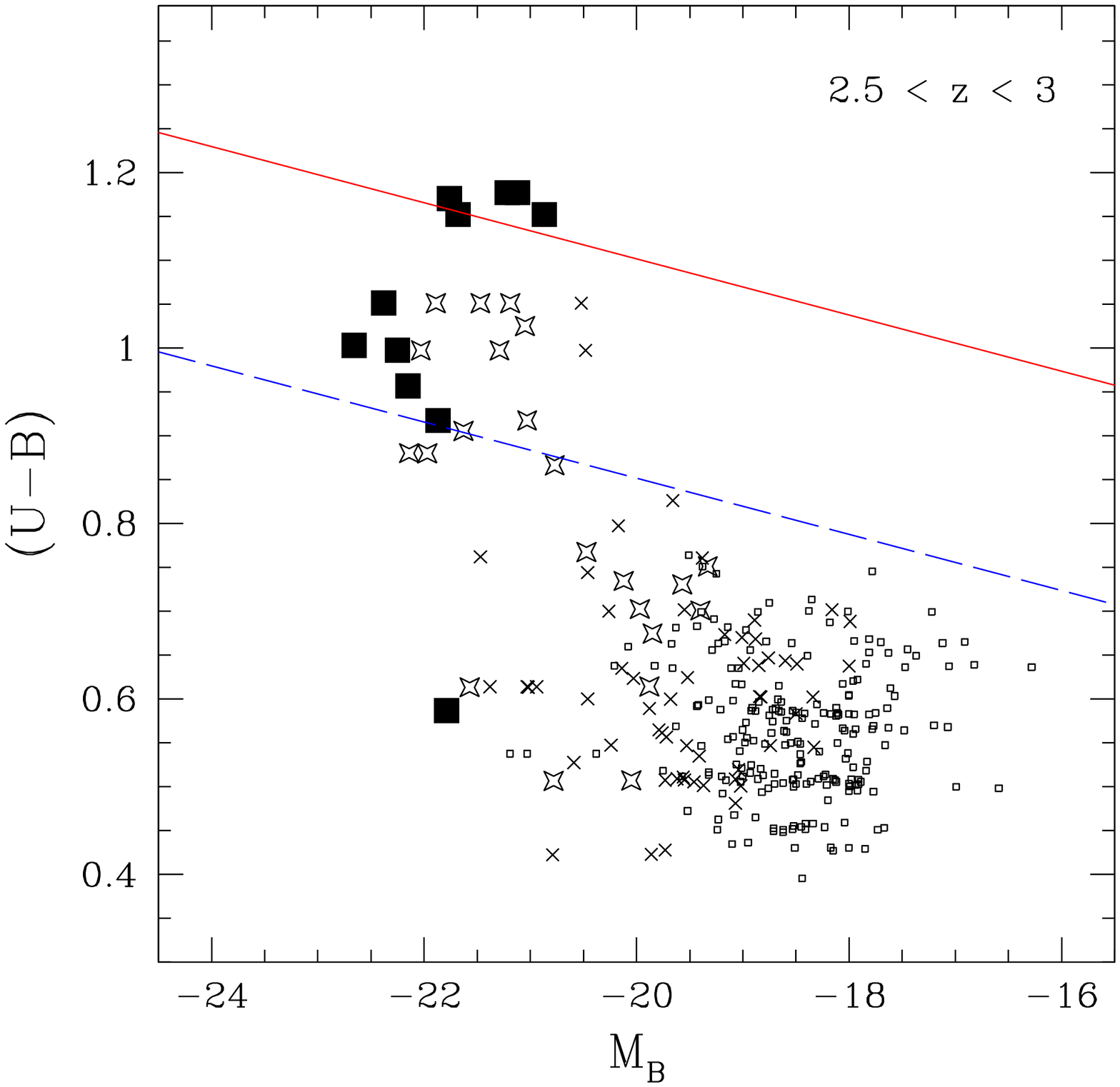}\\
\includegraphics[width=0.28\textwidth]{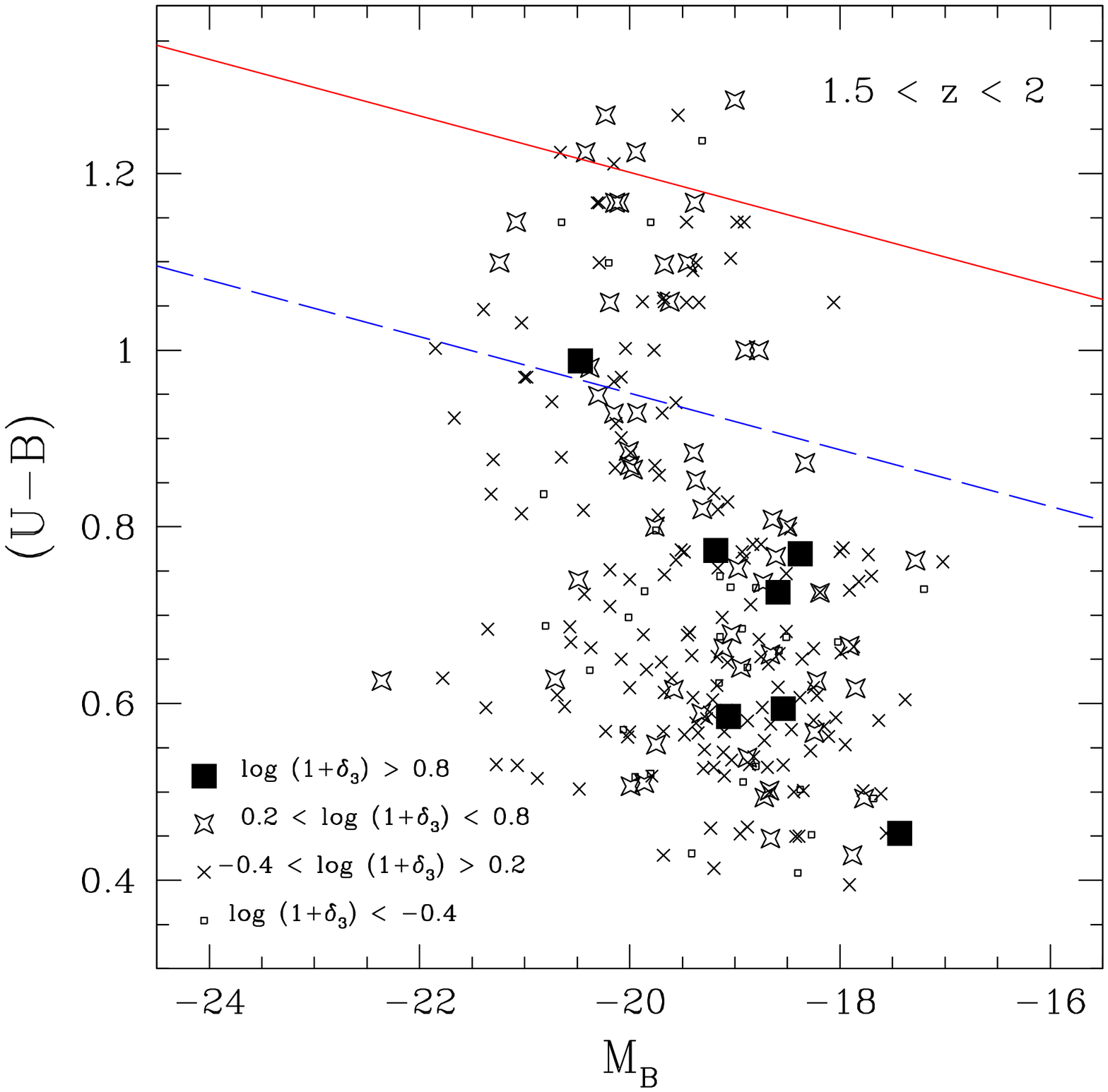}
\includegraphics[width=0.28\textwidth]{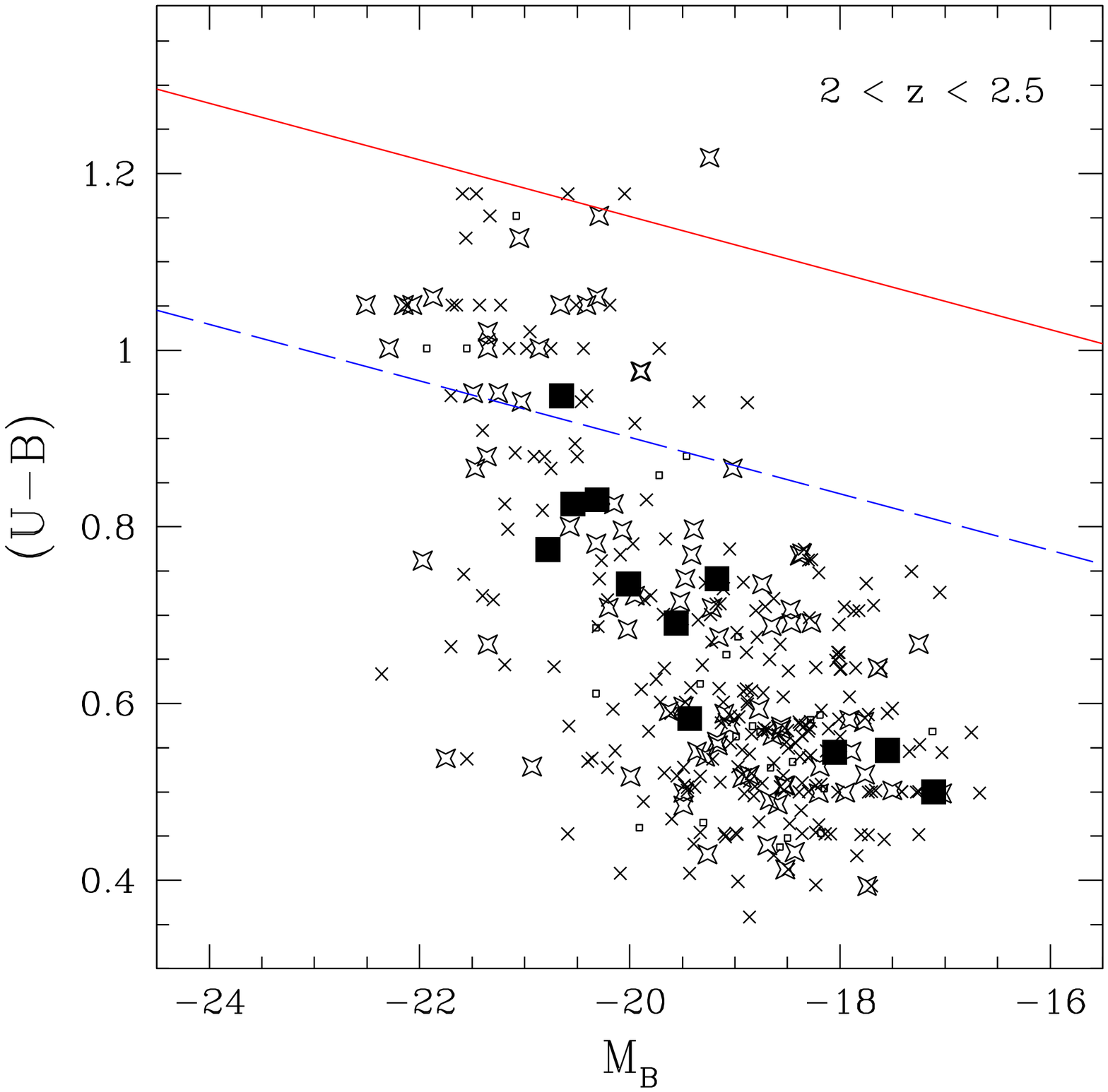}
\includegraphics[width=0.28\textwidth]{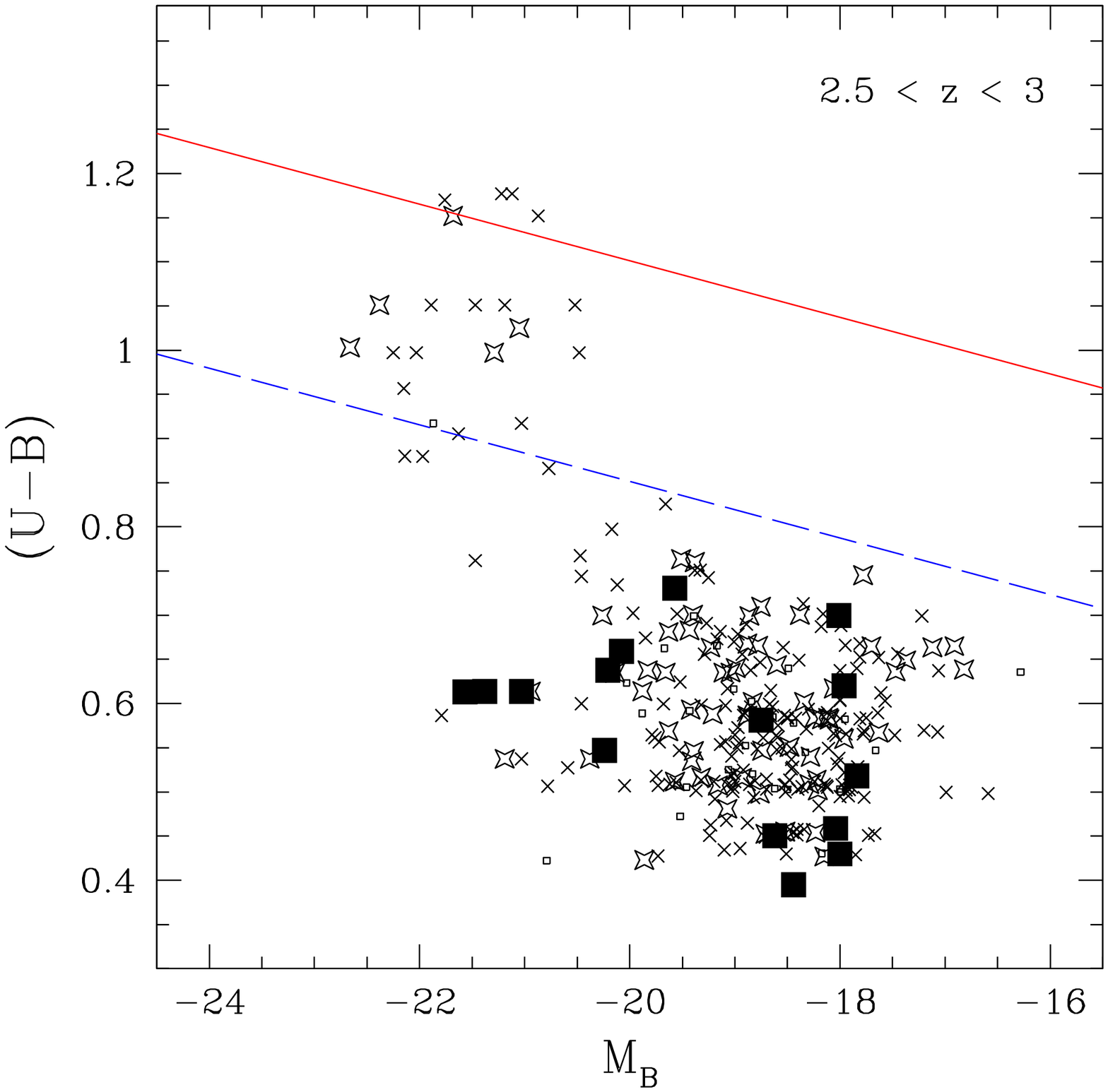}
\caption{Colour-magnitude relation in three redshift ranges, divided in bins of stellar mass (top panels) and local density (bottom panels). The larger symbols correspond to higher stellar mass in the top panels, and to higher overdensity in the bottom panels. The expected location of the red sequence at the respective redshift is shown as red solid line, while the limit to the blue cloud below which a galaxy is considered blue is plotted as blue long-dashed line. \label{fig1}}
\end{center}
\end{figure}

Figure~\ref{fig1} shows the $(U-B)$ - $M_B$ colour-magnitude relation in three different redshift bins: $1.5<z<2$, $2<z<2.5$ and $2.5<z<3$. In the three top panels the sample is split in 4 stellar mass bins with a bin size of 0.5 dex; the symbols are shaped and sized according to the galaxy's stellar mass: larger symbols represent higher stellar mass galaxies. In the three lower panels the sample is split in bins of local density with a bin size of 0.6 dex. Here, larger symbols correspond to galaxies located in higher overdensities. The red solid lines indicate the expected location of the red sequence observed at $z\sim1$, and evolved passively back in time according to the models of \cite{vDF01}. The blue long-dashed lines is the border to the blue cloud, defined as being 0.25 magnitudes bluer than the red sequence.  The different slices in stellar mass are clearly separated in colour-magnitude space. The separation is perpendicular to the red sequence, and the distance from the expected red sequence increases with decreasing stellar mass. The different slices in local density on the other hand are not well separated in colour-magnitude space. The relation between colour and local density is much weaker than the relation between colour and stellar mass. 

\subsection{Colour-density relation and the role of stellar mass}\label{col-dens}

\begin{figure}[t]
\begin{center}
\includegraphics[width=0.44\textwidth]{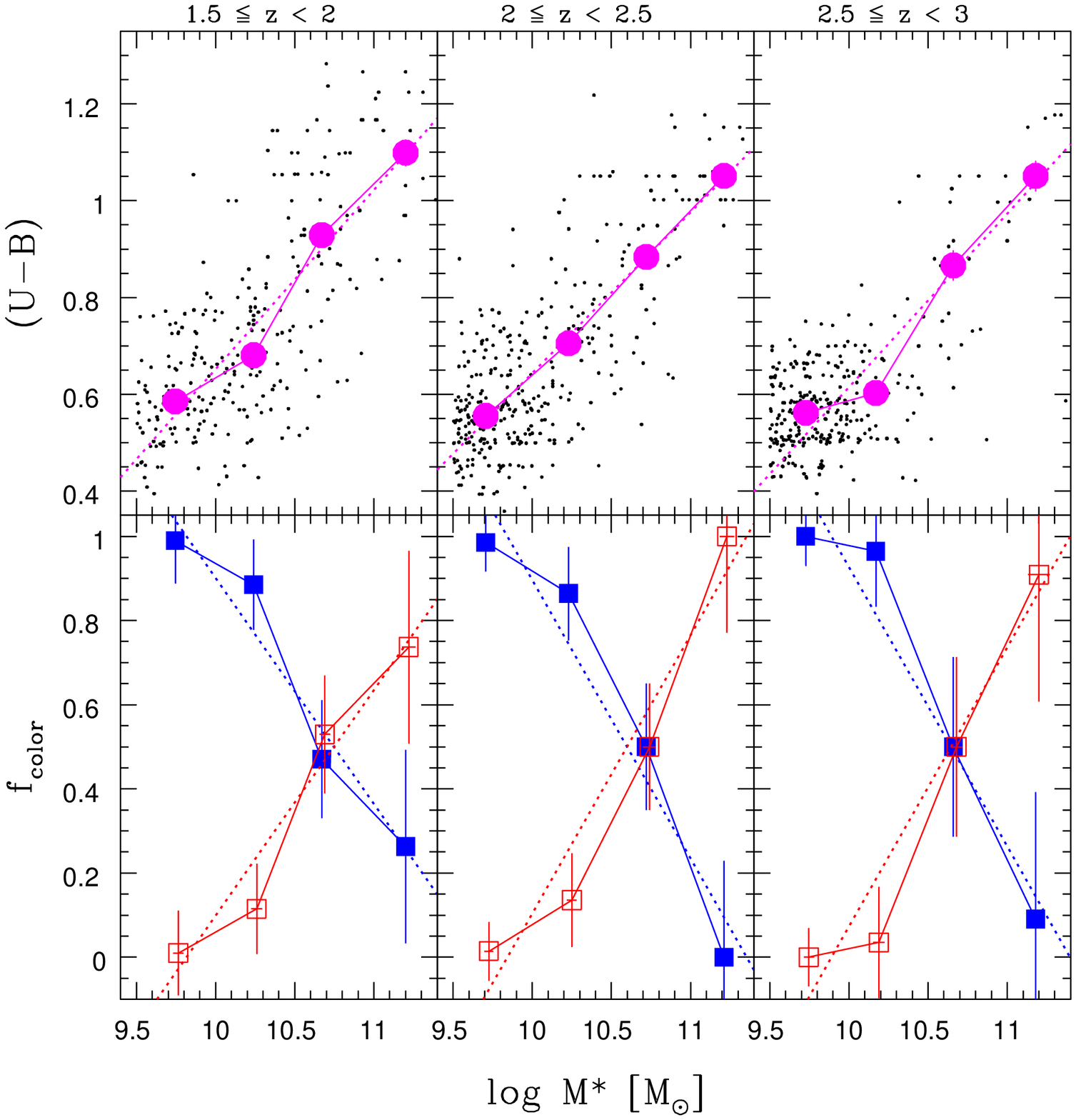}
\includegraphics[width=0.44\textwidth]{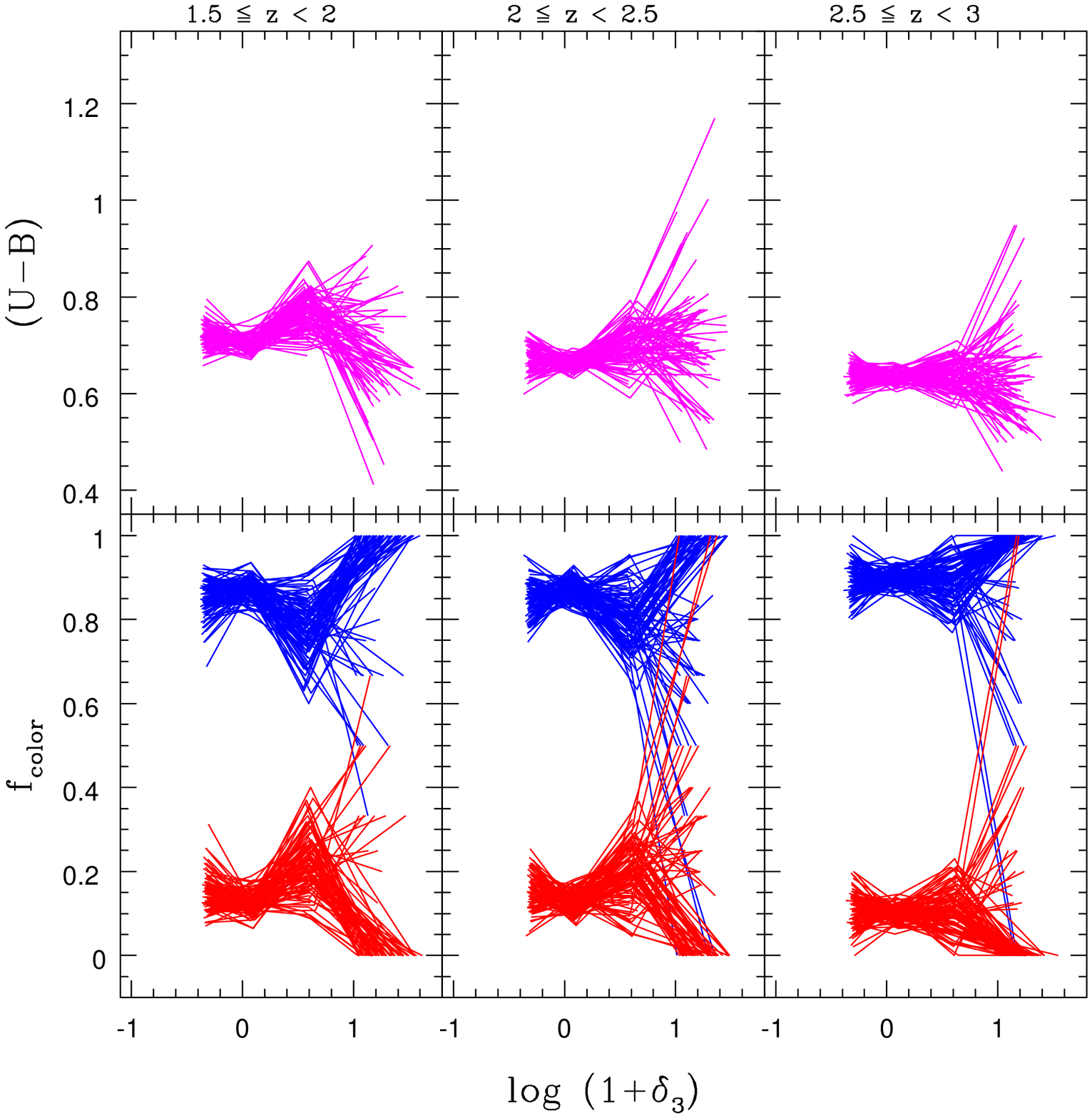}
\caption{$(U-B)$ colour and fraction of red and blue galaxies as a function of stellar mass (left) and local density $\log~(1+\delta_3)$ (right). Each line corresponds to the results of one Monte Carlo Simulation, averaged in bins of stellar mass (left) and local density (right). 
\label{fig2}}
\end{center}
\end{figure}

The left panel of Figure~\ref{fig2}, shows rest-frame $(U-B)$ colour (top) and the fractions of blue and red galaxies (bottom) as a function of stellar mass in three redshift bins. The correlation between $(U-B)$ colour and stellar mass $\log~M_\ast$ is highly significant at all redshifts. The cross-over mass, i.e. the stellar mass at which there is an equal number of red and blue galaxies, stays roughly constant at $\log~M_\ast \sim 10.8$. This is the same cross-over mass found at lower redshifts of $0.4<z<1$ \cite{Gru10}. 

The right panel of Figure~\ref{fig2} shows the relation between $(U-B)$ colour (top row) and the fraction of blue and red galaxies (bottom row) as a function of relative overdensity $\log~(1+\delta_3)$. In this figure we plot the results of the Monte Carlo simulations rather than the original datapoints. Each line shows the average of one Monte Carlo run.  We do not find a significant correlation between colour and overdensity, however, there is a possible trend for a higher fraction of blue galaxies ($\sim 100\%$) at the highest overdensities ($\log~(1+\delta_3) > 0.8$) at all redshifts, where as the blue fraction at intermediate and low densities is around 80-90\%. Note that there is still a 10\% probability that this trend is caused by chance.
%
%We conclude that the colour-density relation at $z>1.5$ is practically not existent or very weak. If a trend with local density persists it must be very minor, with a variation of the average $(U-B)$ colours of less than $\sim$0.1 magnitudes between relative local densities that differ by a factor of 100. There could be an environmental influence on the blue fractions of galaxies in the most extreme overdense environments. This difference in blue fractions amounts to about 10\% and is not detectable at a statistically significant level with the low number of galaxies in our sample located at the highest overdensities (33 galaxies at $\log~(1+\delta_3) > 0.8$ between $1.5 < z < 3$).

\begin{figure}[b]
\sidecaption
%includegraphics[width=0.485\textwidth]{hd_ld_UB_mast_POWIR_NEW_MC.ps}
\includegraphics[width=0.53\textwidth]{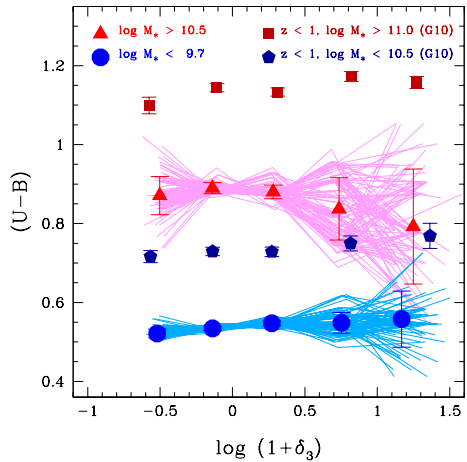}
\caption{Colour density relation in low (blue) and high (red) stellar mass quartiles over the whole redshift range $1.5 < z <3$. The mean and RMS of all Monte Carlo runs in bins of local density are overplotted as red triangles (high quartile) and blue circles (low quartile). The average data points from the sample of \cite{Gru10} at intermediate redshift ($0.4 < z < 1$) are plotted as big red squares (high quartile) and blue pentagons (low quartile) respectively. 
\label{fig3}}
\end{figure}

Figure~\ref{fig3} shows the correlation between colour and local density for galaxies in the low and high stellar mass quartile. A clear colour offset between low- and high-mass galaxies of $\Delta(U-B) \sim 0.2-0.3$ mag is present at all densities. Interestingly, there is a trend that mean colours of galaxies in the high-mass quartile are bluer at higher overdensities ($\log~(1+\delta_3)>0.5$) than at low and average local densities. For galaxies in the low-mass quartile we do not see a strong correlation between colour and local density.

\section{Conclusions}

In this study we investigate the influence of stellar mass and local density on galaxy rest-frame colour at a redshift of $1.5<z<3$. We find the following results:
\begin{itemize}
\item Galaxy colour depends strongly on galaxy stellar mass at all redshifts up to $z\sim3$. After accounting for passive evolution in colour, the colour-stellar mass relation does not evolve with z below that redshift.
\item Massive red galaxies exist up to $z\sim3$, at the expected location of the red sequence in the colour-magnitude diagram.
\item We do not find a correlation between colour and local density, however, galaxies in extremely high over-densities ($>5\times$ over-dense) are bluer than galaxies in average and most under-dense environments.
\item More massive galaxies ($\log M_\ast > 10.5 M_\odot$) at high relative over-densities tend to be on average bluer than massive galaxies at average and low local densities. This is due to a lack of red galaxies at high over-densities.
\end{itemize}
%%% Bibliography

%\input{referenc}
\end{document}